\documentclass[fleqn,10pt]{wlscirep}
\usepackage{todonotes}
\usepackage{threeparttable}
\usepackage{url}
\usepackage{multicol}
\usepackage{adjustbox}
\usepackage{xcolor}
\usepackage{ulem}

\newcommand{\zwRed}[1]{\textcolor{black}{#1}}

\makeatletter
\g@addto@macro\appendix{\setcounter{figure}{0}}
\g@addto@macro\appendix{\setcounter{table}{0}}
\makeatother

\title{\Large\textbf{Urban contact structures for epidemic simulations: Correcting biases in data-driven approaches}
}

\author[1]{Zhanwei Du}
\author[2]{Chao Gao}
\author[3]{Yuan Bai}
\author[3]{Yongjian Yang}
\author[4,*]{Petter Holme}
\affil[1]{Department of Integrative Biology, The University of Texas at Austin, Austin, Texas, 78712, United States of America}
\affil[2]{College of Computer and Information Science \& College of Software, Southwest University, Chongqing 400715, China}
\affil[3]{College of Computer Science and Technology, Jilin University, Changchun, 130012, China}
\affil[4]{Institute of Innovative Research, Tokyo Institute of Technology, Kanagawa 226-8503, Japan}

\affil[*]{holme@cns.pi.titech.ac.jp}

\begin{abstract}
Epidemics are emergent phenomena depending on the epidemiological characteristics of pathogens and the interaction and movement of people. Public transit systems have provided much important information about the movement of people, but there are also other means of transportation (e.g., bicycle and private car), that are invisible to public transit data. This discrepancy can induce a bias in disease models that leads to mispredictions of epidemic growth (e.g., peak prevalence and time). In our study, we aim to advance and compare the epidemic spreading dynamics using public transit trips, in contrast to more accurate estimates of population movement using mobile phones traces. In our study, we simulate epidemic outbreaks in a cohort of two million mobile phone users. We use a metapopulation model incorporating susceptible-infected-recovered dynamics to analyze and compare different effective contract matrices, constructed by the public transit systems and mobile phones respectively, on the process of epidemics. We find that epidemic outbreaks using public transit trips tend to be underestimated in terms of the epidemic spreading dynamics, reaching epidemic peaks weaker and later. This is rooted in a later introduction of new infectious people into uninfected locations. 

\end{abstract}
\begin{document}

\flushbottom
\maketitle
\thispagestyle{empty}
\section*{Introduction}

Despite the advances in medicine and technology, outbreaks of infectious diseases continue to occur, causing substantial losses in both economy and human health. 
For example, in 2014, the Ebola epidemic spreading across Guinea, Liberia, and Sierra Leone, greatly impacted the societies and economies of West Africa~\cite{CDC_cost2017}. In addition to 28,639 cases and 11,316 deaths, the economic loss is estimated to be as high as USD 2.2 billion in 2015 GDP, leaving a declining agricultural production and decreasing cross-border trade~\cite{theworldbank2014,theworldbank2015_Sierra,theworldbank2015_Guinea,theworldbank2015_Liberia}.
To relieve  epidemic risks and reduce socio-economic losses, public health officials seek model-based analyses to inform infectious disease threats (such as the size and time of the outbreak peak) and efficient response strategies to contain and mitigate emerging outbreaks~\cite{giesecke}.

Population movement and connectivity exacerbate epidemic transmission, deciding the fate of the propagation of the epidemics~\cite{dalziel2013human,eubank2004modelling} and the efficacy of public health control measures~\cite{dye2003modeling,riley2003transmission, meyers2005network}. An appropriate example would be the spread of 2009 H1N1 pandemic in Mexico City~\cite{dimitrov2009optimizing,pourbohloul2009initial,merler2011determinants,xia2013identifying}.
Epidemic spreading does not only depend on the transmission mechanisms of the pathogens in question but also population movements.
To get a high-resolution picture of population movements, many technologies have been used to obtain people's trajectories through their digital footprints.
In contrast with individual explosive electronic records available in public transit systems, the position tracking via mobile phones can track nearly total population movements (e.g.\ by foot, bicycle, and public transit) and give a more precise description of population movements during an epidemic~\cite{frias2011agent}. It is however more sensitive from a privacy perspective, which is a reason to keep public transportation data as a source for epidemic modeling.

Fine-grained records of public transit are available in many cities, with public access (e.g., Chicago~\cite{LeeMinjin2015} or Chinese cities such as Chongqing and Shanghai \url{soda.datashanghai.gov.cn}, accessed April 2, 2018).
There is a decades-old line of research in the epidemiological literature to use this data as a proxy for \zwRed{analyzing the features of} population movement~\cite{cooley2011role,zhao2015preliminary,zhang2015including,yashima2014epidemic,saito2013enhancement,colizza2007modeling}. 
However, public transit systems can only track part of movements, dominant in long distance trips. For example, from a study about travel within Shanghai, people tend to travel via public transit for a trip within one hour, but with car for longer journeys~\cite{wang2010car}. Another Chinese study observed that trips within one kilometer were typically by foot, while people used public transportation for longer trips~\cite{Wanghao2014}.

In this work, we aim to advance and compare the epidemic spreading dynamics using population movements captured by different position tracking technologies (e.g.\ mobile phone~\cite{yan2014universal,bengtsson2015using} and public transit~\cite{du2017understanding,sun2015quantifying}).
The epidemic outbreaks are introduced into an urban network of locations, driven by population turnover in metapopulations at the location. The metapopulation dynamics follows a susceptible-infected-recovered model~\cite{colizza2008epidemic}.

\begin{figure}[!tb]
\centering
\includegraphics[width=1\textwidth]{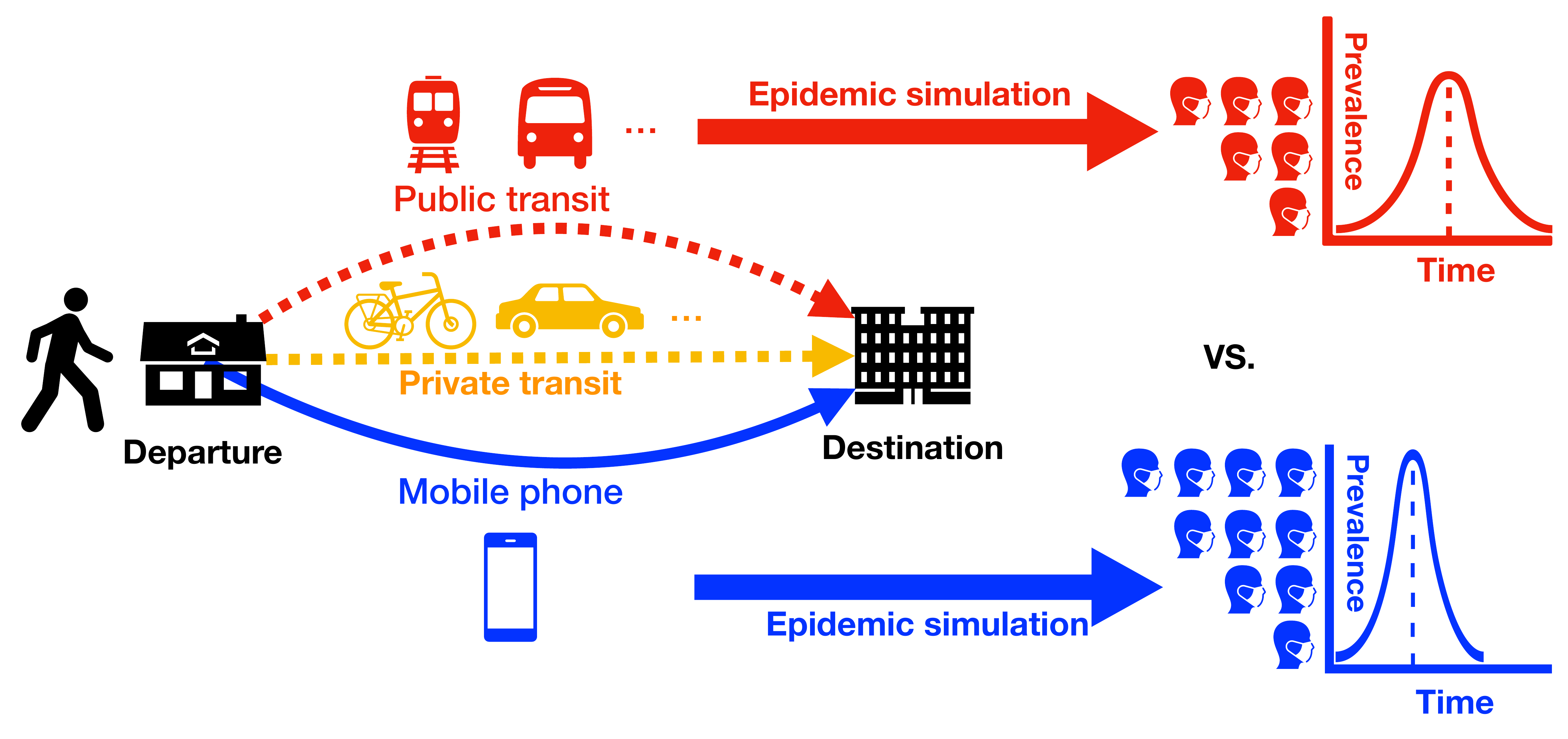}
\caption{{\bf Comparisons of simulated disease outbreaks using public transit or mobile phone data.}
People travel from their origins to destinations using different means of transportation. Using mobile phone data as a basis for simulations could give more accurate results.}
\label{fig:EpidemicFrame}
\end{figure}

The  population movements collected connect over two million mobile phone uses of the eight million inhabitants over one day in Changchun, China in 2017.
As we do not have access to public transportation data from Changchun, we use a model to generate such. This model assumes a gamma distribution of trip distances~\cite{mazloumi2009using}. We track simulated epidemics spreading in epidemiological settings similar to the 2009 Hong Kong H1N1 influenza and 2010 Taiwan varicella epidemics~\cite{yang2016characterizing}, as well as other infectious diseases. We assume the pathogen is introduced in \zwRed{a location probabilistically with respect to its population}
and follow the time evolution of prevalence for the different ways of tracking the populations. In Fig.~\ref{fig:EpidemicFrame}, we show our proposed analysis methods schematically.

\section*{Methods}

In this section, we will introduce the epidemic model, how we deal with movement trips, and prevalence measures in the following subsections. We summarize information about parameters in Table~\ref{Tab:parameters}.

\subsection*{Epidemic model}

The millions of inhabitants of a metropolis make individual-based epidemic models challenging~\cite{MEI201597}. Still one would want to include the location and temporal features of a city in the calculation. One solution, and the one we take, is to represent locations as metapopulations. We base our simulations on the work of Wesolowski \textit{et al.}~\cite{wesolowski2017multinational}. 
We divide the population into three disease compartments: susceptible (individuals don't have but could get the disease), infective (individuals who have the disease and can spread it further) and recovered (individuals who are immune or deceased and cannot get or spread the disease). For each location $j$, the number of individuals in the three compartments at time $t$ are denoted by $S_{t,j}$, $I_{t,j}$ and $R_{t,j}$, respectively. 
In an entirely susceptible population at location $j$, a new outbreak happens with probability: 
\begin{equation}\label{eq:prob}
h(t,j)=\frac{\beta S_{t,j}(1-\exp[-\sum_k m_{j,k}x_{t,k}S_{t,j}] ) }{1+\beta S_{t,j}}
\end{equation}
where $\beta$ represents the transmission rate; 
$m_{j,k}$ is the population flow (i.e., the number of people traveling during one time unit) from location $k$ to location $j$; 
and $x_{t,k}$ is the fraction of the infected population at day $t$ at location $k$
\begin{equation}
x_{t,k} = \frac{I_{t,k}}{N_k} .
\end{equation}
Then we simulate a stochastic process introducing infections into completely susceptible metapopulations. $I_{t+1,j}$ is a Bernoulli random variable with probability  $h(t,j)$.

In locations with infections, the dynamics is deterministic:
\begin{eqnarray}
S_{t+1,j} &=& S_{t,j} -\frac{\beta S_{t,j} I_{t,j}}{N_j}\\ 
I_{t+1,j} &=& I_{t,j} + \frac{\beta S_{t,j} I_{t,j}}{N_j} -\gamma I_{t,j} \\
R_{t+1,j} &=& R_{t,j}+\gamma I_{t,j} 
\end{eqnarray}
where $\gamma$ denotes the recovery rate. 
An important parameter for the discussion is the
basic reproduction number, $R_0$. It is defined as $R_0=\beta/\gamma$ and can be interpreted as the expected number of secondary infections. More specifically, if an infectious individual joins into a susceptible metapopulation. Everyone in such location has an equal probability to contract with other.

\begin{table}
\centering
\caption{Detailed information about parameters in our methods.}
\label{Tab:parameters}
\begin{tabular}{@{}lll@{}}
\toprule
Parameters & Meaning & Range \\ \midrule
$S_{t,j}$ & number of susceptible individuals at time $t$ in location $j$ & $[0,N_j]$ \\
$I_{t,j}$ & number of infective individuals at time $t$ in location $j$ & $[0,N_j]$ \\
$R_{t,j}$ & number of recovered individuals at time $t$ in location $j$ & $[0,N_j]$ \\
$\beta$ & transmission rate & {[}0.5, 15{]} \\
$\gamma $ & recovery rate & {[}1/5,1{]} \\ \midrule
$m_{kj}$ & number of trips from location $j$ to location $k$ in a day & \\
$N_j$ & population of location $j$ & \\
$\mathbf{M}$ & contact matrix using mobile phone trips & \\ \midrule
$\delta$ & public transit average traveling distance & {[}10,50{]} \\
$\mu$ & public transit mode share & 0.3 \\
$k$ & shape parameter in the gamma distribution with respect to $\delta$ & {[}1,50{]} \\
$\theta$ & scale parameter in the gamma distribution with respect to $\delta$ & {[}1,50{]} \\
$d$ & trip distance & \\
$F$ & gamma probability distribution, $\Gamma(k,\theta)$, of public transit trips along with distances $d$ & \\
$\lambda$ & scale parameter to derive the ratio of public transit trips with respect to $\mu$ & \\
$\mathbf{M}_\mu$ & contact matrix using public transit trips & \\ \bottomrule
\end{tabular}
\end{table}

\subsection*{Contact network}

The compartmental model described above is not the only component of an epidemic simulation. We also need to input the contact patterns, which is used to describe who that is in contact with whom. In this section, we describe how we construct the contact matrices that describe the flow of people between the locations within our example city---Changchun, China.

\subsubsection*{Trips inferred by mobile phone data}
The dataset contains the movements of total of 6,305,500 anonymized mobile phone users 
in Changchun City during a one-day period on July 3, 2017 (Monday). 
Fig.~\ref{fig:MoveRatio} shows the temporal number of changing trips in a day.
We use a subset of over 2,035,700 anonymized mobile phone users, who only move within the studied city of Changchun. Changchun is the provincial capital and largest city of the Jilin Province. This city had (in 2010) a total population of 7,674,439 under its jurisdiction and an area of 20,604 $\mathrm{km}^2$.
In total, the data comprise 3,789 cellular base stations (henceforth locations). 

From the movement traces of these users, we construct hourly time series of locations, assuming one individual can only be associated to one location (i.e. the one of longest duration) per hour. At the same time, we construct the contact matrix $\mathbf{M}$, where the entry $m_{kj}$ means the number of trips from location $j$ to location $k$, by aggregating 24 hourly movements of all users.
The population of location $j$, $N_j$, is assumed to be $(m_{jj} + \sum_k m_{jk} -\sum_k m_{kj})/24$, where $m_{jj}$ are the number of local trips within the location $j$.

We summarize some properties of this contact network in Table \ref{Tab:netProperties} and Fig.~\ref{fig:DegreeDis}. The degree $k_i$ of the $i^{th}$ location is the sum of in-degree and out-degree as $(\sum_k m_{ik} + m_{ki})-m_{ii}$. 

\subsubsection*{Public transit trips}

Even though, as mentioned, several metropolitan regions publish detailed data about public transport trips, Changchun does not (at the moment of writing). Note that, there is a sufficient amount of statistics about the city and the intra city travel for us to reconstruct the contact matrix.

We use some different attributes to model public transit trips and extract  the distribution feature of trips: 
First, the average travel distance by public transit, $\delta$, is determined. 
In the city of Changchun, the average speed of public transit is expected to be $15.8$ kilometers per hour in morning and evening rush periods in 2017~\cite{workPg_Changchun}. 
In our analysis, we investigate $\delta$ of three levels to denote traveling trips for $1$ to $3$ hours: low $\delta$ is defined as in the range of 10--20 kilometers, mediate as 30--40 kilometers and high as 50--60 kilometers. 
Next quantity we measure is the public transit mode share.
Another attribute is the public transit mode share, denoted $\mu$---the fraction of public transit trips. According to the survey in Changchun urban area~\cite{Wanghao2014}, the public transit mode share $\mu$ is estimated between 0.3 and 0.4. As thus we set $\mu$ as \zwRed{0.35} in this study.

To generate public transit trips with accurate $\delta$ and $\mu$, we use a realistic probability distribution $F(d)$ of public transit trips of distance $d$.
We assume $F(d)$ follows the gamma distribution---$\Gamma(k,\theta)$. Effectively, $k$ dominates the shape whereas $\theta$ dominates the scale. 
To understand the effect of $\delta$ on $F(d)$, we use gamma distributions with mean value as $\delta$. We scan the integer values of the parameter space of $k$ from two to fifty and $\theta$ from one to fifty.
For example, consider the low $\delta$, then we choose all pairs of $k$ and $\theta$, giving a mean of $\Gamma(k,\theta)$ between 10 and 20. 
Totally, there are 29 pairs of $k,\theta$-values for low $\delta$, 37 pairs for mediate $\delta$, and 34 pairs for high $\delta$. 
For each $\Gamma(k,\theta)$, $\lambda$ is introduced explicitly to scale the probability when deriving the fraction of public transit trips as $\mu$. 
Finally, a trip is labeled as public transit with the probability of $\lambda \Gamma(k,\theta)$.
By aggregating public transit trips, we can construct the contact matrix $\mathrm{M}_\mu$ of public transit trips, the same process as that of mobile phone trips.

\subsection*{Simulation setup and analysis}

As epidemic outbreaks emerge, their transmission dynamics can vary substantially across the geographical locations of the initial infection in large populations\cite{dalziel2013human,eubank2004modelling}, which have been used to model epidemics in urban settings \cite{cooley2011role,dalziel2013human,herrera2016disease}.
We make 100 runs for averages. Each run starting from a location chosen with a probability proportional to its population.
The outbreaks are characterized by three parameters (the transmission rate $\beta$, the recovery $\gamma$, and the the average traveling distance $\delta$): 
 $\beta$  indicates the probability of a susceptible individual infected in a contact by an infectious individual. $\gamma$'s inverse gives an expected duration of the infectious state. Finally, $\delta$ parameterizes the public transit behavior of individuals.
Simulations with parameter settings of infectious diseases are studied:
2009 Hong Kong H1N1 influenza ($\beta=0.50$ and $\gamma=1/3$) and 2010 Taiwan varicella ($\beta=1.55$ and $\gamma=1/5$)~\cite{yang2016characterizing}, as well as a series of hypothetic infectious diseases.
We simulate the hypothetic diseases by three levels of transmission rates (low $\beta=2$, mediate $\beta=5$, high $\beta=15$ and $\gamma=1 $)~\cite{wesolowski2017multinational}.
For example of a set of parameter settings (i.e.,  $\beta=1/2$, $\gamma=1/3$ and low $\delta$), 
there are 29 pairs of $k$ and $\theta$ with respect to low $\delta$. For each pair of $k$ and $\theta$, we run stochastic simulations \zwRed{30} times using sampled public transit trips. 
With $100 \times (1+29+37+34)\times 30\times(6+3)=2,727,000$ simulations, we assess these simulated prevalence time-series with respect to the following four comparisons:
\begin{itemize}
\item Early warning: The time lag, as $t_\mathrm{1\%}^{MPT}-t_\mathrm{1\%}^{PTT}$, between the prevalence with public transit trips (PTT) reaching 1\% prevalence at time $t_\mathrm{1\%}^{PTT}$ and the prevalence with mobile phone trips (MPT) reaching 1\% at time $t_\mathrm{1\%}^{MPT}$prevalence~\cite{herrera2016disease,thompson2006epidemiology,rath2003automated}.
\item Peak timing: The time lag, as $t_\mathrm{peak}^{MPT}-t_\mathrm{peak}^{PTT}$, between the prevalence with public transit trips reaching its epidemic peak at time $t_\mathrm{peak}^{PTT}$ and the prevalence with mobile phone trips reaching its epidemic peak at time at time $t_\mathrm{peak}^{PTT}$~\cite{herrera2016disease}.
\item Peak magnitude: The ratio, as $r_\mathrm{peak}^{PTT}/r_\mathrm{peak}^{MPT}$, of peak prevalence $r_\mathrm{peak}^{PTT}$ with public transit trips and that $r_\mathrm{peak}^{MPT}$ with mobile phone trips~\cite{herrera2016disease}.
\item Situational awareness: The complement of the normalized mean absolute error (MAE) between prevalences of simulations using public transit trips and that using mobile phone trips, minimized over possible lags~\cite{herrera2016disease}, given by:
\begin{equation}
1-\min_{\lambda} \frac{\sum_t \left | x_t - y_{t+\lambda} \right |} {\sum_t \left | x_t + y_{t+\lambda}\right |}
\end{equation}
Here, $x_t$ and $y_t$ are the prevalence in simulations using public transit trips and that using mobile phone trips at time $t$, respectively; $\lambda$ is the time lag.
\end{itemize}

To track the epidemic growth rate of new infections in previous susceptible locations, we consider the dynamic percentage of infected locations over time:
\begin{itemize}
\item Timing of \#\% locations infected: The time lag as $l_\mathrm{1\%}^{MPT}-l_\mathrm{1\%}^{PTT}$, between the prevalence with public transit reaching \#\% locations infected at time $l_\mathrm{1\%}^{PTT}$ and that $l_\mathrm{1\%}^{MPT}$ with mobile phones.
\end{itemize}

\section*{Results}

\subsection*{Evaluation of prevalence time-series}

\begin{figure}[!ht]
\centering
\includegraphics[width=1\textwidth]{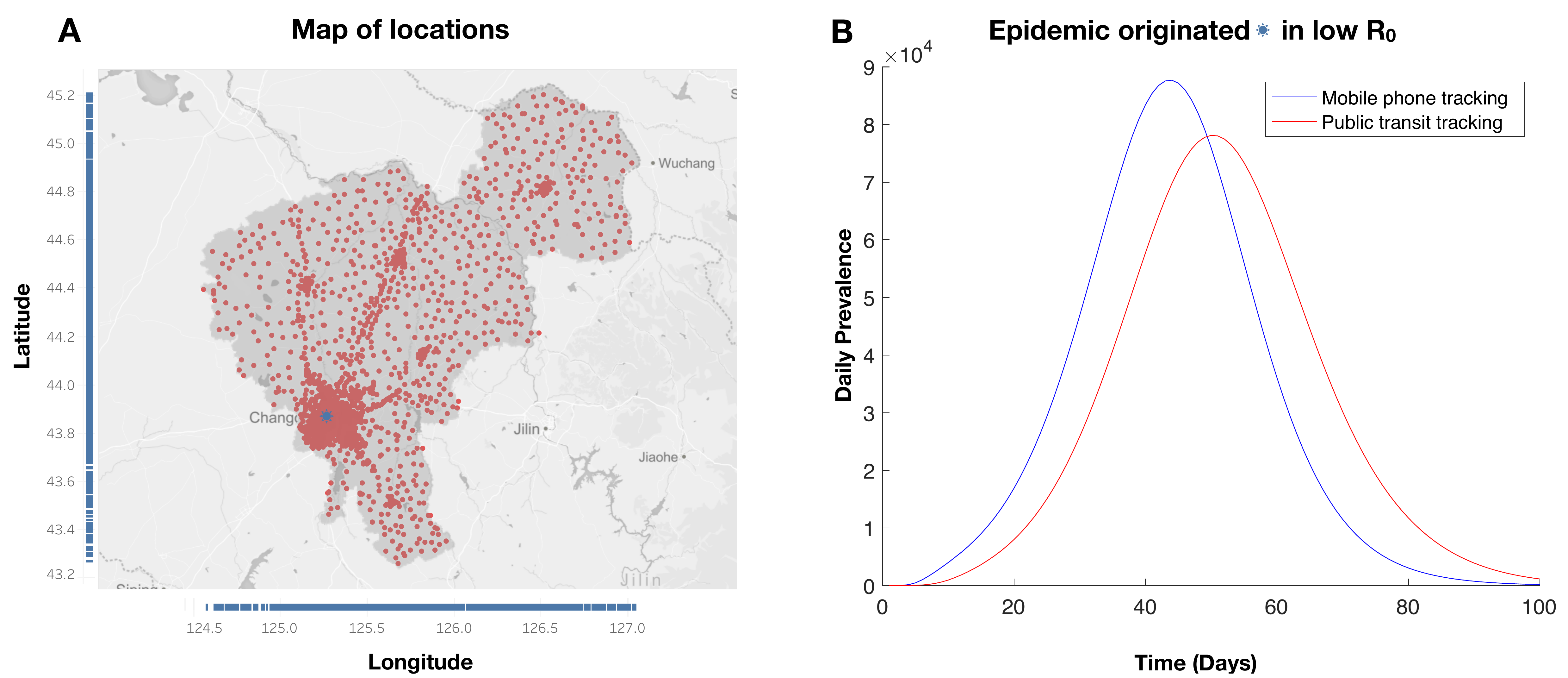}
\caption{{\bf Map of 3,789 locations in our metapopulation model and an example of typical epidemic curves with parameters from the 2009 Hong Kong H1N1 Influenza Pandemic (low $R_0=1.5$).}
In panel A,
The horizontal (vertical) axis denotes the latitudes (longitude) of locations. 
We initialize the outbreak by infecting the most populous location (star).
The movement pattern of public transit is featured by the mediate public transit average traveling distance (30--40 kilometers) and the gamma distribution ($k$ as two and $\theta$ as 16). 
In panel B, we show an example of a prevalence curve. Simulations based on population movements from mobile phones peak at nearly the 40$^\text{th}$ day, while that from public transit at 50$^\text{th}$ day.
}
\label{fig:Example}
\end{figure}

As mentioned, we scan the $\beta$-$\gamma$ parameter space. A typical epidemic starts with an initial infection at a random location. 
Except the outbreak size, the epidemic growth can be additionally different.
Considering an epidemic as shown in Fig.~\ref{fig:Example}, prevalences using public transit trips are increasing with time, with the peak at the 50$^\text{th}$ day and end at the 170$^\text{th}$ day, in contrast, that using mobile phone trips peak at the 40$^\text{th}$ day.

The epidemic spread is closely tracked in epidemiological settings of 2009 Hong Kong H1N1 influenza ($\beta=0.5$ and $\gamma=1/3$) and 2010 Taiwan varicella ($\beta=1.55$ and $\gamma=1/5$)~\cite{yang2016characterizing}, as well as a series of synthetic, hypothetical infections. 
Infections of disease outbreaks in urban settings may be detected in one day~\cite{wesolowski2017multinational}. We further track the epidemic spread of synthetic infectious diseases over transmission rates (low $\beta=2$, mediate $\beta=5$, and high $\beta=15$).

\begin{figure}[!ht]
\centering
\includegraphics[width=1\textwidth]{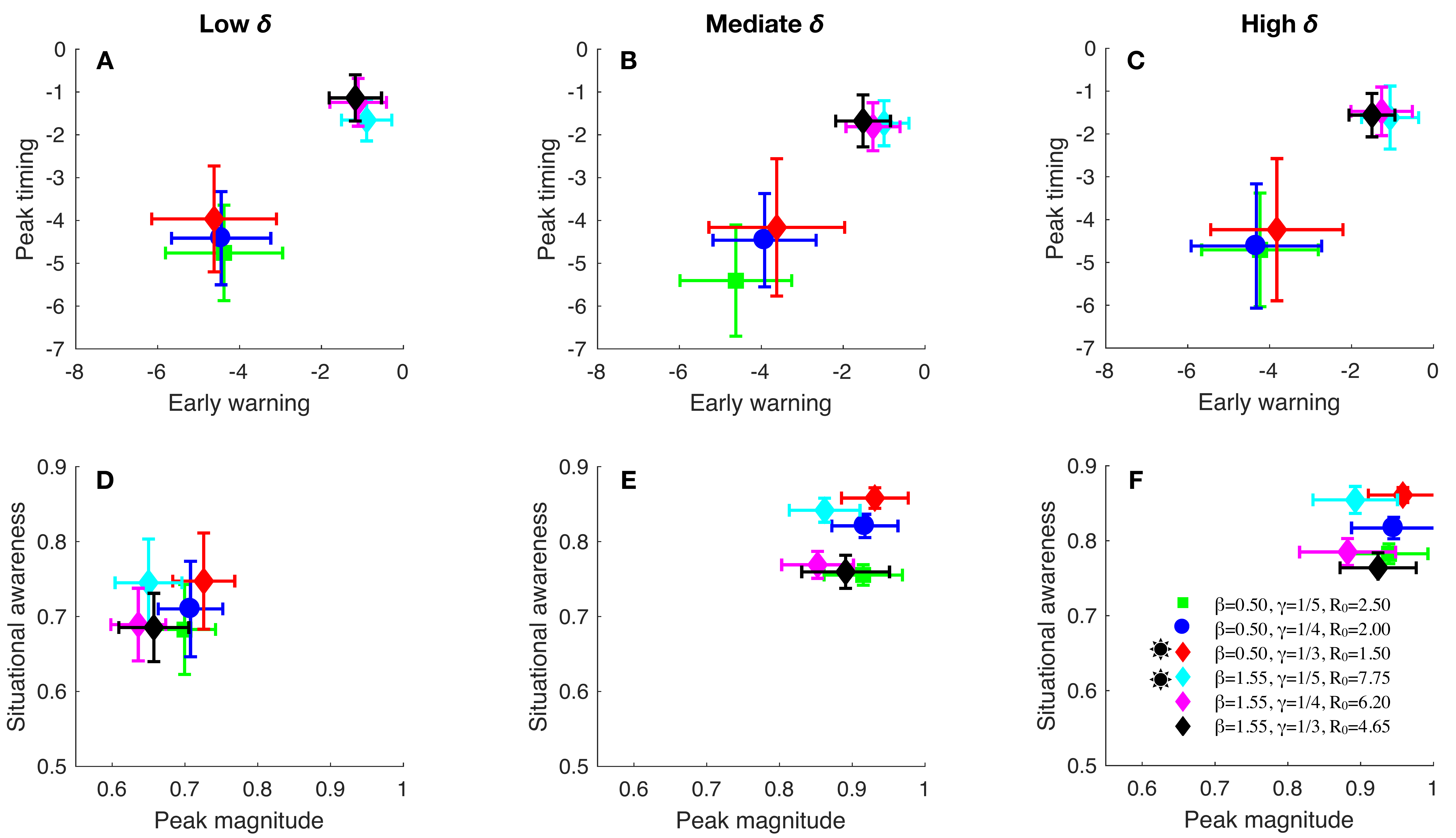}
\caption{{\bf Performance of two real infectious diseases with respect to early warning and peak timing (graphs A, B, and C), and achieving peak magnitude and situational awareness (graphs D, E, and F).}
Points and error bars indicate the mean and standard deviation in performance over simulations, respectively. We choose parameter values to reflect the 2009 Hong Kong H1N1 influenza ($\beta=0.5$ and $\gamma=1/3$) and 2010 Taiwan varicella ($\beta=1.55$ and $\gamma=1/5$)~\cite{yang2016characterizing} marked by stars in legend. We also include simulations for other parameter values representing other infectious diseases.}
\label{fig:Real}
\end{figure}

\begin{figure}[!ht]
\centering
\includegraphics[width=1\textwidth]{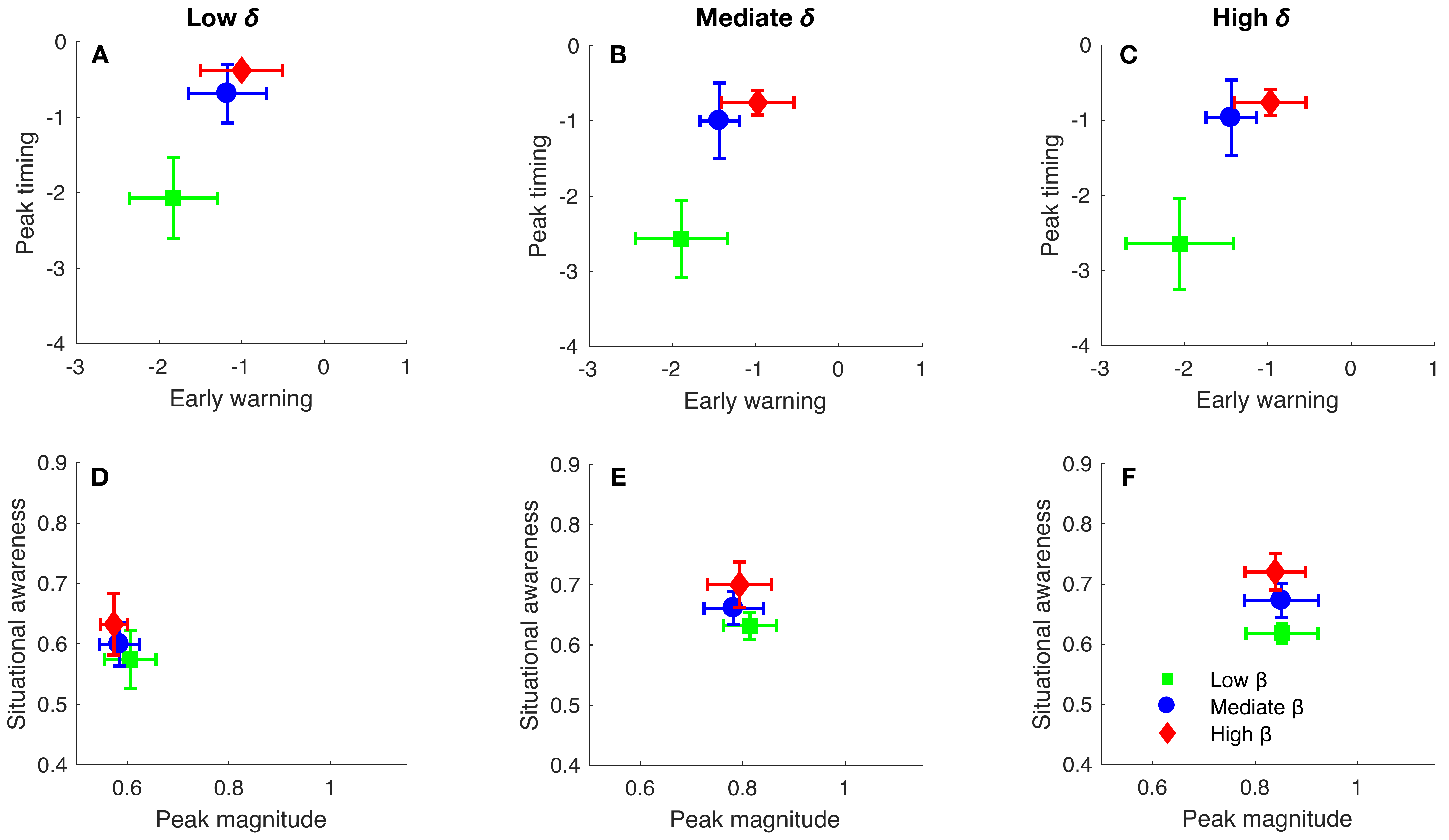}
\caption{{\bf Performance of hypothetical infectious diseases with respect to early warning and peak timing (graphs A, B, and C), and achieving peak magnitude and situational awareness (graphs D, E, and F).}
Points and error bars indicate the mean and standard deviation in performance over simulations, respectively. 
We simulate hypothetical diseases with three levels of transmission rates (low $\beta=2$, mediate $\beta=5$, and high $\beta=15$) ~\cite{wesolowski2017multinational}.
In the simulated epidemic of each $R_0$ and each average traveling distance $\delta$ (low, mediate, and high), negative early warning and peak timing indicate the late arrival of peak prevalence in simulations using public transit trips.}
\label{fig:Syn}
\end{figure}

Figs.~\ref{fig:Real} and \ref{fig:Syn} show the deviations of the simulations based on transit data from the mobile phone based reference. 
The consistently negative times, shows that both the peak times and early warning times always happens later in the simulations based on public transit data.
The small peak magnitudes and situational awarenesses also shows a discrepancy between the outbreaks estimated from the public transit trips and mobile phone trips.
 In general, one can regard the bias as coming from that the simulations based on the public transit trips are more strongly affected by both $R_0$ and the average traveling distance $\delta$.
For example of low $\delta$,  when $R_0 = 7.75$, the peak time of prevalence using public transit data is on average at the 17$^\text{th}$ day, and that using mobile phone at the  the 15$^\text{th}$ day. As thus, the peak timing, as the deviation of their peak times, is on average two days. 
When $R_0 = 1.5$, the peak time of prevalence using public transit data is on average at the 48$^\text{th}$ day, and that using mobile phone at the  the 44$^\text{th}$ day. The peak timing  increases to an average of four days (Fig.~\ref{fig:Real}A, \ref{fig:Real}B and \ref{fig:Real}C).

\begin{figure}[!ht]
\centering
\includegraphics[width=1\textwidth]{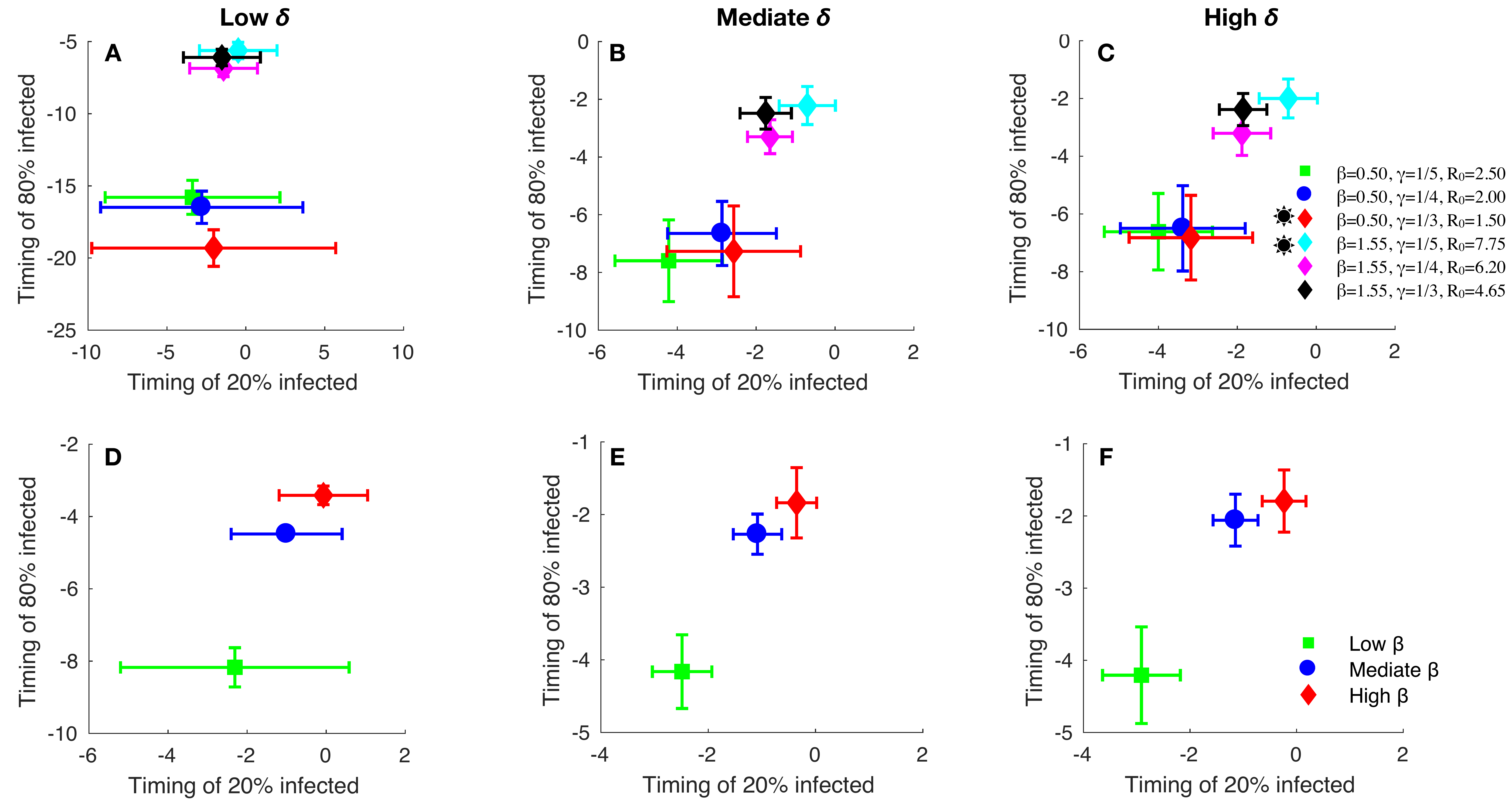}
\caption{{\bf Epidemic growths of real (graphs A, B, and C) and synthetic (graphs D, E, and F) infectious diseases with respect to timing of 80\% and 20\% locations new infected.}
Points and error bars indicate mean and standard deviation in performance over simulations, respectively. 
We closely track the timing of \#\% locations new infected for real infectious diseases (Fig.~\ref{fig:Real}) and synthetic infectious diseases (Fig.~\ref{fig:Syn}).
In the simulated epidemic of each $R_0$ and each average traveling distance $\delta$ (low, mediate, and high), 
The negative times indicates a predicted slower epidemic spreading using public transit trips data in terms of incidence.}
\label{fig:TimingInfect}
\end{figure}


\subsection*{Theoretical network analysis}

To better understand our observations, we investigate the difference in population flow as predicted by public transit and mobile phone data, respectively. Suppose that an entirely infected location $i$ of size $N_i$ is connected to an entirely susceptible location $j$ of size $N_j$. 
Considering this simple scenario, we can follow the network analysis of Ref.~\citen{caillaud2013epidemiological} to understand this effect of contact matrices on epidemics. The probability $P_{ji}$, that the disease is transmitted between an infected individual from the $i^\text{th}$ location go to the $j^\text{th}$ location, is given by:
\begin{equation}
P_{ji} = 1-\lim_{dt \rightarrow 0} \left(1 - \left[\frac{\beta m_{ji}n_{j}}{n_{i}}\right] dt\right)^{\gamma/dt} = 1- \exp\left(\frac{-\beta \gamma m_{ji}n_{j}}{n_{i} }\right)
\end{equation}
Where $m_{ji}$ denotes the movement flow as the number of trips from location $i$ to location $j$ per time---the entry of contact matrix $\mathbf{M}$. 
Thus the probability $\Theta_{ji}$ that at least one infected individual of location $i$ due to location $j$ is given: 
\begin{equation}
\Theta_{ji} = 1- \exp\left(-\beta \gamma m_{ji}n_{j}\right) = \lim_{m_{ji}\rightarrow 0} \beta \gamma m_{ji}n_{j} =\lim_{m_{ji}\rightarrow 0} R_0 m_{ji}n_{j} 
\end{equation}
When $m_{ji}$ is small, the probability $\Theta_{ji} $ can be linear with the term of $ R_0 m_{ji}$.
As thus, $\Theta_{ji} $, as the probability of new infections introduced in an entirely susceptible location, correlates positively with $R_0$ and movement flow $m_{ji}$. 

For an infectious disease with fixed $R_0$, comparing with the contact matrix $\mathbf{M}$ of mobile phone trips, the contact matrix $\mathbf{M}_\mu$ of public transit trips, with the entry of $m_{ji}$, is scaled by $\mu$ and has different probability distributions. 
For example of mediate  $\delta$ as 30--40 kilometers~(Fig.\ref{fig:Distribution}), 95\% trips  tracked by mobile phone are covered  by a  range from 0 to 153 kilometers. In contrast, 95\% public transit data has a more narrow range (i.e. from 0 to 84 kilometers), only about half of the mobile phone data.
These differences contribute to the disparity of probability $\Theta_{ji}$ to infect new locations in epidemic outbreaks.

Furthermore, we evaluate the kind deviation of epidemic prediction between the two types of movement data by considering the expected first time of  \#\% locations being infected. 
In Fig.~\ref{fig:TimingInfect}, we track the spread of realistic infectious diseases (Fig.~\ref{fig:Real}) and hypothetic infectious diseases (Fig.~\ref{fig:Syn}). Negative times shows the slower epidemic growths  using public transit trips. 
In general, times are decreasing with increasing $R_0$ and public transit average traveling distance $\delta$. 
For example, at the beginning, simulations with public transit trips arrive around when 20\% locations new infected later or earlier. With continued epidemics, these simulations will arrive the time of 80\% locations new infected later totally.

\section*{Conclusions}

In computational epidemiology, a common proxy for population flow within cities is public transit data. In this work, we compared such an approach with arguably more realistic data for population movement obtained from mobile phone locations. Simulating outbreaks in several (realistic and hypothetical) scenarios, we found deviations depending on the type of population movement data we input. In the simulations based on public transit data, the peaks of  outbreaks tended to arrive later and be weaker. For example, for H1N1 to spread over a medium traveling distance (30--40 kilometers), prevalences using public transit have can have five days delay and be 8\% less of the peak when using cell phone data. 

We trace the observed effect to a delayed spreading of new infections in totally susceptible locations. New infections driven by public transit data tend to arrive late at an entirely susceptible population, especially for locations with a long distance,  due to the significantly reducing number of trips and appearance probability of long trips.
We conclude that deviations of epidemic prevalence using public transit decreased with increasing $R_0$ and public transit average traveling distance $\delta$.
This conclusion, in general, could prove useful to public health policymakers, especially when there are only public transit information, and can guide future tracking strategies to infer population movements. It also implies that some earlier literature needs to be re-evaluated. 
For example of  the  epidemic risk relied on the public transit data in Tokyo~\cite{yashima2014epidemic}, with the public transit mode share as 57\%~\cite{tokyo2012}, the risk that a single infection causes Tokyo epidemic may be underestimated.  With a low transmission rate $\beta$ as $10^{-4}$ and a high work population as $10^4$,  the previous low probability of urban epidemic can be middle or even high, and the peak time can arrive  weeks earlier than the estimated two months.

Our work agrees with Ref.~\citen{MEI201597} in that we need a holistic picture of movements within a city. Future studies along this line can either increase the precision in the population density and flow modeling~\cite{bart}, introduce demographic data~\cite{chen2010effect,MEI201597}, or scale up the modeling to the entire humanity~\cite{balcan2010modeling,colizza2007modeling}.

\appendix

\section*{Supporting information}
ZW would like to acknowledge funding from the Models of Infectious Disease Agent Study (MIDAS) program grant number U01 GM087719.
This work was also funded by the National Natural Science Foundation of China (NSFC) (Grant No. 61772230 and 61402379).
The funders had no role in study design, data collection and analysis, decision to publish, or preparation of the manuscript.

\begin{figure}
\centering
\includegraphics[width=0.5\textwidth]{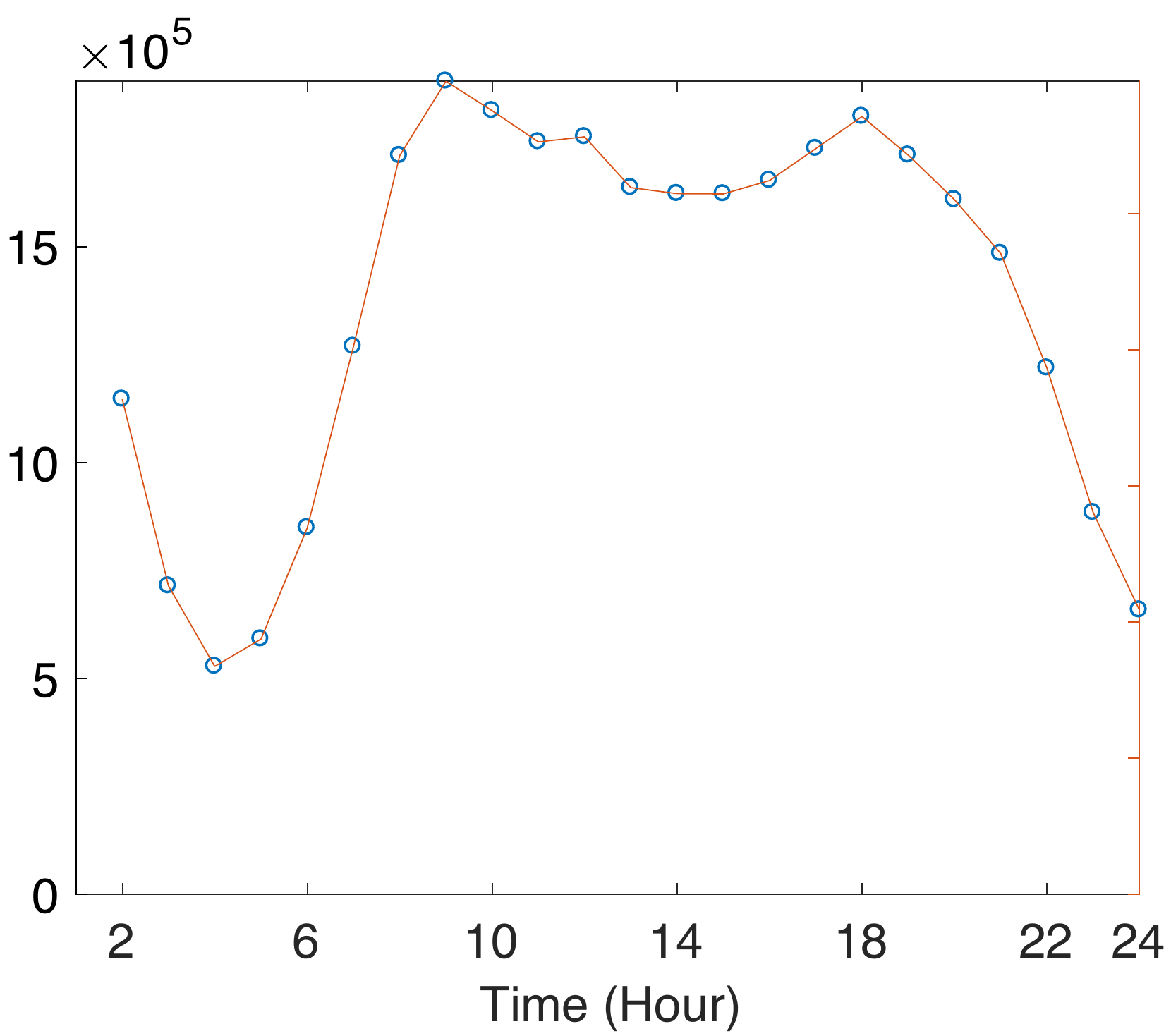}
\caption{{\bf Number of trips across locations in the studied day. X axis denotes 23 hours in a day starting from 2 am.}
There are two peaks in a day, i.e., morning peak staring from 9 am and evening peak starting from 6 pm.}
\label{fig:MoveRatio} 
\end{figure}

\begin{table}
\centering
\caption{
{\bf Properties of the contact network with respect to mobile phone trips.} Number $N$ of different locations generated $E$ edges during $T$ time steps. $\Delta t$ represents time units, $\langle k\rangle$ represents average degree, $M$ represents modularity of the contact network, estimated by the method of Louvain~\cite{PhysRevE.70.066111} via 46 detected communities (a set of locations densely interconnected, and with few connections to locations outside).}
\begin{tabular}{llllll}
\hline
$N$ & $E$ & $T$ & $\langle k\rangle$ & $M$ \\ \midrule
3946 & 1099422 & 1d & 15285 & 0.89 \\ \bottomrule
\end{tabular}
\label{Tab:netProperties}
\end{table}

\begin{figure}
\centering
\includegraphics[width=0.55\textwidth]{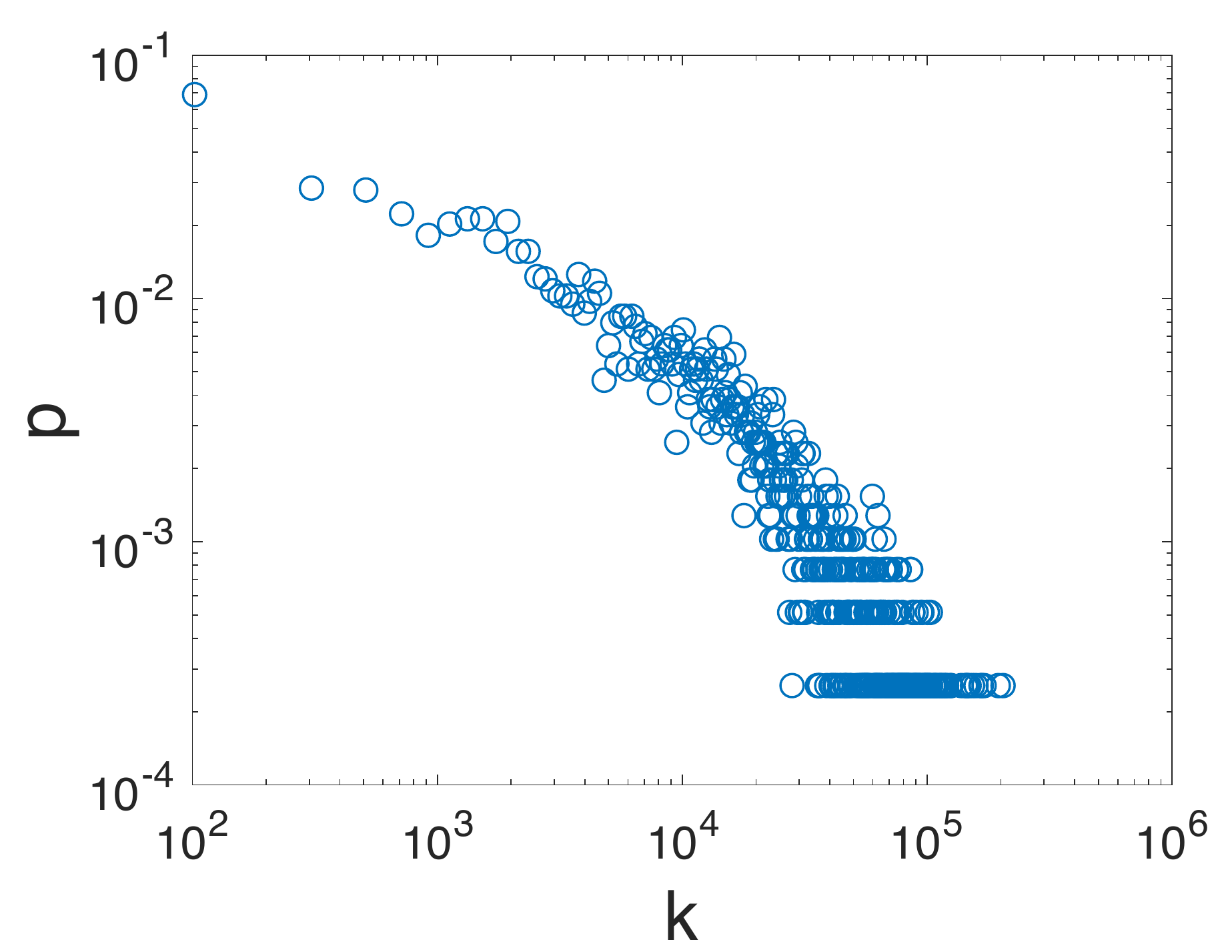}
\caption{{\bf The statistical characteristics of the contact network with respect to mobile phone trips.}
We plot the probability density $p$ as a function of degree $k$.}
\label{fig:DegreeDis}
\end{figure}

\begin{figure}
\centering
\includegraphics[width=1\textwidth]{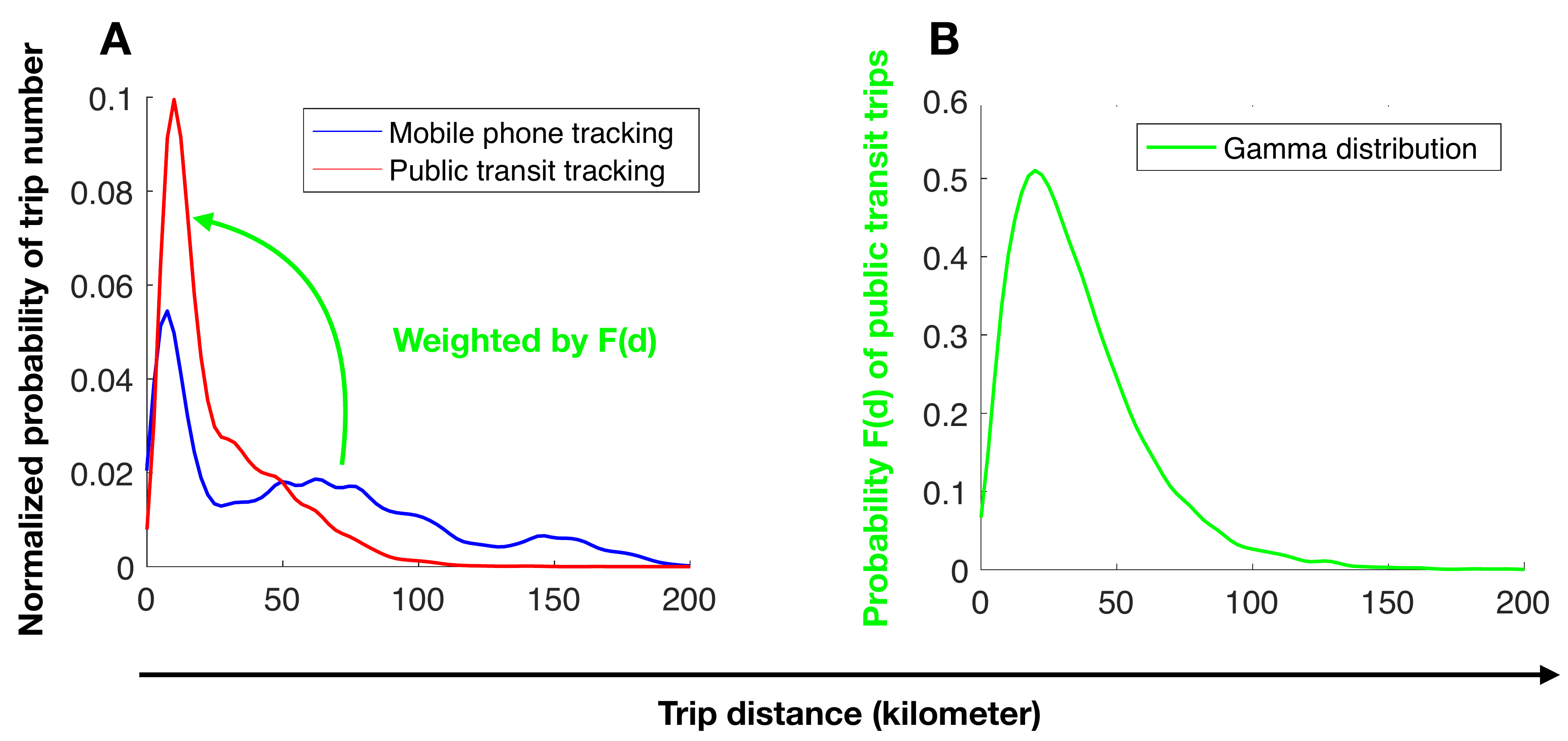}
\caption{{\bf Example of public transit trips originated from mobile phones.}
Here the public transit average traveling distance,  $\delta$, is mediate as 30--40 kilometers.
The left figure shows the normalized probability distributions of trip number tracked by mobile phones (red curve) and public transit (blue curve). The blue curve is weighted with respect to the probability $F(d)$ of public transit trips following the gamma distribution ($k$ as 2 and $\theta$ as 16) in the right figure.}
\label{fig:Distribution}
\end{figure}

\bibliography{new}

\end{document}